\documentclass{article}
\usepackage[final]{graphics}
\usepackage{bbm}
\usepackage{amsbsy}
\usepackage{amssymb,amsfonts}

\DeclareFontFamily{U}{bbm}{}
\DeclareFontShape{U}{bbm}{m}{n}{%
<5><6><7><8><9><12>gen*bbm%
<10>bbm12%
<10.95>bbm10%
<14.4>bbm12%
<17.28><20.74><24.88>bbm17%
}{}

\newcommand\imb{\boldsymbol}

\begin{document}
\date{January 1999}

\bibliographystyle{unsrt}

\begin{titlepage}

\title{\begin{center}
{\Huge LAPTH}
\end{center}
\vspace{5 mm}
\hrule
\vspace{10mm}
\bf{A new approach\\
 for the vertical part of the contour\\
 in thermal field theories}}
\author{Fran\c cois~Gelis\footnote{email: gelis@lapp.in2p3.fr}}
\maketitle

\vskip 5mm
\begin{center}
Laboratoire de Physique Th\'eorique LAPTH,\\
B.~P.~110,  F--74941 Annecy-le-Vieux Cedex, France
\end{center}

\vskip 1cm
\begin{abstract}
  A lot of work has been devoted in the past to understand the role of
  vertical branch of the time path in thermal field theories, and in
  particular to see how to deal with it in the real-time formalism.

  Unlike what is commonly believed, I emphasize on the fact that the
  vertical part of the path contributes to real-time Green's
  functions, and I prove that this contribution is taken into account
  simply by the substitution $n(\omega_{\boldsymbol k})\to n(|k_o|)$
  in the real time Feynman rules. This new proof is based on very
  simple algebraic properties of the contour integration.

\end{abstract}
\vskip 4mm \centerline{\hfill LAPTH--713/99, hep-ph/9901263\hglue 2cm}

\vfill
\thispagestyle{empty}
\end{titlepage}

\section{Introduction}
\begin{figure}[htbp]
\centerline{\resizebox*{5cm}{!}{\includegraphics{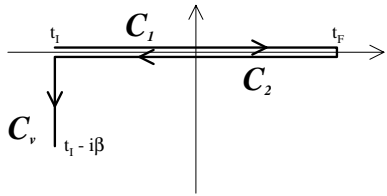}}}
\caption{Time path in the closed time path formalism.\label{fig:path}}
\end{figure}

When deriving the matrix Feynman rules of the closed time path (CTP
in the following) formalism, an intriguing problem is to understand
how one can reach a $2$ components matrix formulation from the path
represented on figure \ref{fig:path}.
Indeed, it is widely accepted that each component of the matrix
formalism corresponds to one of the two horizontal branches ${\cal
  C}_1$ and ${\cal C}_2$ of the path
\cite{KobesK1,KobesSW1,LandsW1,FurnsS1,Marin1,NiemiS1}. In this
picture, there is no room for the vertical part ${\cal C}_v$.
Therefore, most of the derivations of the matrix formalism found in
the literature got rid in some way of the vertical part of the path.
Usually, one invokes the limits $t_{_{I}}\to -\infty$ and
$t_{_{F}}\to+\infty$, in conjunction with an {\it ad hoc\/} choice of
the asymptotic properties of the source $j(x)$ coupled to the field in
the generating functional.

This derivation seems highly artificial, as one can judge by the
numerous attempts \cite{Niega1,Evans6,EvansP1} to find other, less
{\it ad hoc\/}, justifications. Among them, the most ``revolutionary''
approach was that of Niegawa \cite{Niega1} who rejected the hypothesis
according to which the vertical part of the time path does not
contributes to Green's functions in the real time formalism.  Instead,
he argued that it does contribute in some cases, and this contribution
can be taken into account by the so called ``$n(|k^o|)$
prescription''. Although correct in his statement, his proof is
controversial since he still makes use of the artificial limits
$t_{_{I}}\to -\infty$ and $t_{_{F}}\to+\infty$.

In a previous paper \cite{Gelis1}, I attempted to show this result
without using these limits at all. Indeed, I started by showing that
(i) the vertical part of the path contributes in general, (ii) the
time integrations involved in the calculation of Feynman diagrams in
time coordinates give a result which is totally independent of the
time $t_{_{I}}$ and $t_{_{F}}$. As a consequence, all the arguments
based upon specific limits for $t_{_{I}}$ and $t_{_{F}}$ were suspect
since nothing nontrivial could occur when taking these limits. After
that, I showed in the case of self-energy insertions between
propagators that the contribution from the vertical part is precisely
the term that corresponds to the difference between the $n(|k^o|)$
prescription and the $n(\omega_{\imb k})$ prescription. This proof was
quite intricate, mainly because it involved dealing with delicate
products of distributions, and seemed also to leave open the
possibility for the vertical part to contribute in many other cases.

After that, the intricacies of the products of distributions were
elegantly avoided by Le Bellac and Mabilat \cite{BellaM1,Mabil1}, who
introduced a regularization of the propagators, which had the main
property of preserving the Kubo Martin Schwinger (KMS in the following)
boundary conditions, as well as the holomorphy of the
propagators.\footnote{Other justifications of the matrix formalism
  used regularizations as well, like \cite{LandsW1,EvansP1}. But, in
  these papers, the regularization scheme had the effect to break KMS
  and to make the contribution of the vertical part artificially
  vanish.  Then, one introduces by hand the ``$n(|k^o|)$
  prescription'' in order to reinforce KMS, and it happens that this
  prescription is precisely what was needed to take into account the
  contribution of the vertical part, as we shall see later. A
  consistent justification of the matrix formalism should never need
  to ``reinforce KMS'' since it should never use intermediate steps
  that break KMS.} In their proof, the necessity of using $|k^o|$ as
the argument of statistical factors appeared quite naturally, but in a
way which was not obviously related to the vertical part of the path.

My purpose in the present paper is to present an alternative proof of
this result, in a way which avoids all the intricacies of the
multiplication of distributions, while being more complete than
\cite{Gelis1} since all the situations in which the vertical part of
the path can contribute are clearly identified. The part of the proof
dealing with self-energy insertions, which was nontrivial in
\cite{Gelis1}, is now quite straightforward thanks to the use of
simple algebraic properties of the contour integration.

The structure of this paper is as follows. In section
\ref{sec:origin}, I start by recalling the origin of the vertical part
of the path, and its precise role for the consistent perturbative
expansion of a theory of quantum fields in thermal equilibrium.  Then,
in section \ref{sec:tTOE}, I explain why performing the Fourier
transform to go from the time variable to the energy variable is more
complicated at finite temperature than it is at zero temperature. In
this section, I also derive the matrix formalism in a naive (and
incorrect) way, assuming first for the sake of simplicity that the
vertical part of the path does not contribute.

Section \ref{sec:localisation} is devoted to a detailed study of
the circumstances in which the vertical part of the path contributes.
It is shown that it can contribute only in two simple cases: vacuum
diagrams and self-energy insertions.

In section \ref{sec:effect}, I study the effect of the vertical
part in the case of self-energy insertions,\footnote{The necessity of
  modified Feynman rules in order to calculate vacuum diagrams in
  the real time formalism is known for a long time. Justifications can
  be found in \cite{Evans1,Evans2,Gelis1}.} and show how the matrix
Feynman rules must be modified to generate properly the contribution
of the vertical part. This proof starts by the almost trivial case of
repeated concatenation of free propagators, which is then generalized
to the case of general self-energy insertions by simple algebraic
arguments.  Finally, the last section is devoted to concluding
remarks. Some technical details are relegated into two appendices.

\section{Origin of the vertical part}
\label{sec:origin}

In \cite{Gelis1}, I derived the perturbative expansion of a thermal
field theory in time coordinates by using the canonical approach in
order to make more explicit the role of the vertical part of the path.
I will just summarize here the main points of this derivation. When
doing this perturbative expansion, the main difference with respect to
the zero temperature situation is related to the fact that the
parameter of the expansion (the coupling constant of the theory)
appears not only in the dynamics of the fields via their evolution
equation, but also in the averaging procedure itself via the density
operator $e^{-\beta H}$ ($\beta\equiv 1/T$, $k_{_{B}}=1$). Indeed, the
Hamiltonian $H$ contains the coupling constant. One then sees easily
that the two horizontal branches are necessary in order to expand in
powers of the coupling constant the time evolution of the fields. But
in order to have a consistent perturbative expansion (in particular to
preserve thermal equilibrium order by order in the coupling constant),
one needs also to expand in powers of the coupling constant the
density operator itself. This is done easily thanks to the following
formula \cite{RammeS1}\footnote{The reason why such a formula is
  possible is related to the analogy between the canonical density
  operator and an evolution operator. The role of the vertical part of
  the path seems to have remained unnoticed by particle physicists,
  who usually derived the perturbative expansion of thermal field
  theories by functional methods based on the Feynman-Kac formula
  \cite{LandsW1}.}
\begin{equation}
e^{-\beta H}=e^{-\beta H_o}\; {\rm T}_c\,\exp\, i\,
\int_{{\cal C}_v\times{\mathbbm R}^3}
{\cal L}_{\rm in}(\phi_{\rm in}(x))\,d^4x\; ,
\end{equation}
where $H_o$ is the free part of the Hamiltonian, ${\cal C}_v$ is a
path in the complex time plane going from $t_{_{I}}$ to
$t_{_{I}}-i\beta$, ${\cal L}_{\rm in}$ is the interaction part of the
Lagrangian density, and $\phi_{\rm in}$ is the field in the
interaction picture (i.e. a free field). With this formula, it is now
obvious that the perturbative expansion of the density operator itself
made possible by the addition of the vertical part ${\cal C}_v$ to the
previous two horizontal branches.

The physical meaning of the vertical part ${\cal C}_v$ is now quite
clear: this piece of the contour is needed because the interaction
modifies the equilibrium density operator. Therefore, it is likely
that this vertical part is crucial for the consistency of the
perturbative expansion, and that arguments suggesting that it can
simply be dropped are wrong.\footnote{We are now in a position to
  understand why enforcing by hand KMS and taking into account the
  vertical part can be related: without the vertical part, the
  perturbative expansion would be inconsistent because the density
  operator would not be expanded in powers of the coupling constant.
  In other words, statistical equilibrium, i.e. KMS, would be broken.}

\section{From time to energy -\\
 First approach to the RTF}
\label{sec:tTOE}
\subsection{From time to energy}
At this stage, we have definite Feynman rules to calculate
perturbatively a Green's function in time coordinates: at each vertex,
one must integrate over time along the whole path ${\cal C}\equiv{\cal
  C}_1\cup{\cal C}_2\cup{\cal C}_v$. As at zero temperature, the
problem is that these Feynman rules are not very convenient for
practical calculations. One usually prefers to work in the Fourier
space with the conjugate variables $(k^o,{\imb k})$.  Since in the
sector of spatial variables, everything is similar to the zero
temperature case, going from position to 3-momentum is trivial and
works exactly in the same way as at $T=0$ (in the following, I assume
that the transformation ${\imb x}\to{\imb k}$ has already been
performed, and I do not write explicitly the spatial variables).

Problems arise when one tries to go from time to energy. Indeed, the
property behind the usefulness of the Fourier transform is the
relation existing between the Fourier transform (FT in the following)
and the convolution product. More precisely, given two 2-point
functions $f(x^o_1,x^o_2)$ and $g(x^o_1,x^o_2)$, one expects the FT to
satisfy the identity
\begin{equation}
FT(f*g)(k^o_1,k^o_2)=[FT(f)(k^o_1,k^o_2)][FT(g)(k^o_1,k^o_2)]\; .
\label{eq:FT-conv}
\end{equation}
The problem comes from the fact that the relevant convolution product
at finite temperature is defined by an integration along the path
$\cal C$ instead of the real axis ${\mathbbm R}$:
\begin{equation}
(f*g)(x^o_1,x^o_2)\equiv\int_{{\cal C}}
dy^o\,f(x^o_1,y^o)g(y^o,x^o_2)\; .
\label{eq:conv-def}
\end{equation}
Obviously, the usual definition of the Fourier transform cannot
accommodate the relations of Eq.~(\ref{eq:conv-def}) and
Eq.~(\ref{eq:FT-conv}). This definition should be modified in order to
make these relations compatible. A first solution that I will not
develop here is provided by the so called imaginary time formalism,
which can be seen in this context as a work-around for the above
problem.  More precisely, one makes use of the $-i\beta$-periodicity
properties of thermal Green's functions in order to expand them in
Fourier series, the Fourier modes (called Matsubara frequencies in
this context) being imaginary since the period is imaginary.

\subsection{Naive approach to the RTF}
Another solution is provided by the matrix formulation (often called
real time formalism when the context makes obvious the fact that we
are in the Fourier space). For each $n$-point function, one defines
$2^n$ distinct Fourier transforms labelled by $n$ superscripts $a_i=1$
or $2$, via the relations:\footnote{I am implicitly assuming that
  ${\cal C}_{1,2}$ are extended from $-\infty$ to $+\infty$ in this
  definition of the Fourier transforms, in order to make them as close
  as possible to the usual one. Nevertheless, it should be emphasized
  that this limit has no effect as far as the contribution of the
  vertical part is concerned, since the integrand $G(x_1,\cdots,x_n)$
  is totally independent upon the times $t_{_{I}}$ and $t_{_{F}}$ (see
  appendix \ref{sec:appA}). Therefore, we still have to find how to
  deal with the contribution of ${\cal C}_v$ in this matrix
  formalism.}
\begin{equation}
G^{\{a_i\}}(k_1,\cdots,k_n)\equiv
\left[
\prod\limits_{i=1}^{n}\int_{{\cal C}_{a_i}\times{\mathbbm R}^3}d^4x_i\;
e^{ik_i\cdot x_i}
\right]\;
G(x_1,\cdots,x_n)\; .
\label{eq:FT-def}
\end{equation}

Then, one would like to have Feynman rules enabling a direct
calculation of these new Green's functions, without going through the
stage of the function in time coordinates. As a first approach, let us
first assume that the vertical part ${\cal C}_v$ does not contribute
to the calculation of the function $G(x_1,\cdots,x_n)$. This
hypothesis has been at the basis of most of the attempts to derive the
matrix formalism, and the focus has mainly been on findings arguments
to justify it. In the present paper, I will first derive the Feynman
rules of the real time formalism in situations where the vertical part
does not contributes. If we assume that the vertical part of the path
does not contribute in the convolution $P(x_1,x_2)\equiv
(F*G)(x_1,x_2)$, then we have obviously in terms of the previously
defined Fourier transforms:
\begin{equation}
P^{ab}=F^{a1}G^{1b}-F^{a2}G^{2b}=(F\tau_3 G)^{ab}\; ,
\label{eq:feyn-naive}
\end{equation}
where the Pauli matrix $\tau_3\equiv{\rm Diag}(1,-1)$ deals with the
minus sign associated to type $2$ indices.  This relation can be seen
as a particular form of Eq.~(\ref{eq:FT-conv}), the product of the
right hand side being a matrix product. Therefore, we see that
transforming 2-point functions into $2\times 2$ matrices enables to
generalize the usual relationship between the Fourier transform and
the convolution product to the thermal case.

Going on along this line, we would of course obtain the standard
matrix formulation for the real time formalism in Fourier space.
Nevertheless, this justification is valid only for situations in which
the vertical branch does not contribute. At this point, the standard
way has been to try to get rid of the vertical part. Instead of that,
I will determine precisely the situations in which it contributes, and
show that its contribution can be included in the matrix formalism by
a minor modification of its Feynman rules.

\section{Diagrams in which\\
 the vertical part contributes}
\label{sec:localisation}
\subsection{Example}
A simple example showing that the vertical part can contribute to the
result of a path integration is provided by the convolution of two
bare propagators. Such a calculation would appear for instance in the
insertion of a mass term. This calculation has been done explicitly in
\cite{Gelis1} and shows the following features:

(i) The vertical part is mandatory in order to have a result invariant
under time translation.

(ii) The vertical part enables to get rid of the $t_{_{I}}$ dependence
that would show up in the result if one where using only ${\cal
  C}_1\cup{\cal C}_2$ (the validity of this result is quite general,
see appendix \ref{sec:appA}).

(iii) A $t_{_{I}}$-independent, invariant under time translation,
contribution of the vertical part of the path is left in the result.

The existence of such an explicit example definitively rules out the
justifications based on the initial hypothesis that the vertical part
does not contribute.

\subsection{Generic contour integration}
It is convenient to work with the mixed coordinates $(x^o,{\imb k})$
in which the bare propagator has the following explicit expression:
\begin{equation}
G_o(x^o,y^o;{\imb k})={1\over{2\omega_{\imb k}}}\sum\limits_{s=\pm}
G_{o,s}^{\omega_{\imb k}}(x^o,y^o)\; ,
\label{eq:bare-prop}
\end{equation}
with
\begin{equation}
G_{o,s}^E(x^o,y^o)\equiv e^{-isE(y^o-x^o)}
\left[
\theta_c(s(y^o-x^o))+n_{_{B}}(E)
\right]
\label{eq:G-o-E}
\end{equation}
and
\begin{equation}
\omega_{\imb k}\equiv \sqrt{{\imb k}^2+m^2}\qquad\qquad 
n_{_{B}}(E)\equiv{1\over{e^{\beta E}-1}}\; .
\end{equation}
Because of the structure of this bare propagator, it is {\it a
  priori\/} obvious that every time integration can be reduced to
integrals of the following type:
\begin{equation}
I_{\cal C}(\Sigma)\equiv \int_{\cal C}dx^o\,e^{-i\Sigma x^o}f(x^o,\Sigma)\; .
\end{equation}
In the above integral, $\Sigma$ is a linear combination of the
on-shell energies $\omega_{{\imb k}_i}$ corresponding to the various
legs (internal as well as external) of the diagram, with coefficients
$0$, $1$ or $-1$, while the function $f(\cdot)$ is a product of
factors like $\theta_c(\pm(x^o-x^o_i))+n_{_{B}}(\omega_{{\imb k}_i})$.
This function is therefore piece-wise constant along the path $\cal
C$. Moreover, the KMS boundary condition is such that the integrand
takes equal values at both ends of the path:\footnote{In situations
  where fermions with chemical potential are present in the theory,
  this result remains true because the fermions always come in pairs
  at vertices and because charges are conserved at each vertex.}
\begin{equation}
e^{-i\Sigma t_{_{I}}}f(t_{_{I}},\Sigma)=e^{-i\Sigma(t_{_{I}}-i\beta)}
f(t_{_{I}}-i\beta,\Sigma)\; .
\end{equation}

I want now to show that the object $I_{\cal C}(\Sigma)$ receives a
contribution of the vertical part ${\cal C}_v$ if and only if
$\Sigma=0$. Let us first assume that $\Sigma\not=0$. Therefore, an
integration by parts gives immediately:
\begin{equation}
I_{\cal C}(\Sigma)={1\over{i\Sigma}}\int_{\cal C}dx^o\,e^{-i\Sigma x^o}
\;{{\partial f(x^o,\Sigma)}\over{\partial x^o}}\; .
\label{eq:int-parts}
\end{equation}
Then, since the function $f()$ is piece-wise constant, its derivative
is a discrete sum of Dirac's distributions $\delta_c()$. Therefore,
the generic structure of the above integral is a sum like
\begin{equation}
I_{\cal C}(\Sigma)={1\over{i\Sigma}}\sum\limits_{i}c_i e^{-i\Sigma x^o_i}\; ,
\end{equation}
where the $c_i$ are coefficients we don't need to make more explicit
($c_i$ is the value at the point $x^o=x^o_i$ of the coefficient in
front of $\delta(x^o-x^o_i)$ in $\partial f/\partial x^o$)  and the
$x_i^o$ are the times at which the value of $f(x^o,{\Sigma})$ changes.
Now, in order to see if there is in the above result a contribution
which is specific to the vertical part of the path, let us calculate
the same Feynman diagram using only ${\cal C}_1\cup{\cal C}_2$. This
means that $x^o$, as well as all the $x^o_i$ are now restricted to the
horizontal branches of the time path.  For the integral $I(\Sigma)$,
the result would be the same sum restricted to those times $x^o_i$
that are on the horizontal branches:
\begin{equation}
I_{{\cal C}_1\cup{\cal C}_2}(\Sigma)=
{1\over{i\Sigma}}\sum\limits_{\{i|x^o_i\in {\cal C}_1\cup{\cal C}_2\}}
c_i e^{-i\Sigma x^o_i}\; .
\end{equation}
But, by definition of the calculation based on only ${\cal
  C}_1\cup{\cal C}_2$, all the other times $x^o_i$ are also on ${\cal
  C}_1\cup{\cal C}_2$, so that the ``restricted'' sum contains in fact
all the terms of the full sum, with the same coefficients $c_i$.
Therefore:
\begin{equation}
{\rm if\ \ } \Sigma\not=0,\qquad I_{\cal C}(\Sigma)=
I_{{\cal C}_1\cup{\cal C}_2}(\Sigma)\; ,
\end{equation}
and there is no contribution specific to ${\cal C}_v$ in this case.
In other words, all the contour integrals give the same result whether
they appear in the calculation with the full path or in the
calculation with only the horizontal branches, if $\Sigma\not=0$.

Let us now consider the case where $\Sigma=0$. The integration by
parts gives now
\begin{equation}
I_{\cal C}(\Sigma)={-i\beta}f(t_{_{I}},0)-\int_{\cal C}dx^o\,
{{\partial f(x^o,0)}\over{\partial x^o}}\; .
\end{equation}
By the same arguments as before, we can show that there is no
contribution specific to the vertical part in the second term. But now
the factor $-i\beta$ in the first term comes from the difference
$t_{_{I}}-(t_{_{I}}-i\beta)$ of the two extremities of the time path.
Therefore, this term would vanish if we were dropping the vertical
part. From that, we conclude that this first term is a contribution
from the vertical part.\footnote{We see that the contribution of the
  vertical part is not a continuous function of $\Sigma$.
  Nevertheless, the total contribution is a continuous function of
  $\Sigma$.}

\subsection{Localization of the contribution of ${\cal C}_v$}
The condition $\Sigma=0$ necessary to have a contribution of the
vertical part is a constraint on the 3-momenta (both internal and
external) of the diagram. But not all the situations where $\Sigma=0$
lead to a contribution of the vertical part at the very end of the
calculation. Indeed, since the function $I_{\cal C}(\Sigma)$ is
continuous at $\Sigma=0$, we won't have a contribution of ${\cal C}_v$
at the end if the condition $\Sigma=0$ defines a sub-manifold of zero
measure in the space accessible to 3-momenta (taking into account the
constraints provided by 3-momentum conservation). This is in fact the
generic case.

There are only two distinct situations in which the condition
$\Sigma=0$ does not reduce the accessible space more than the
3-momentum conservation does. The first of these two cases correspond
to vacuum diagrams (diagrams without external legs) for which the last
time integration has always $\Sigma=0$ because of the invariance under
time translation (because a function of a single time must be a
constant if invariance under time translation holds). The fact that
the last time integration plays a particular role in such a diagram is
at the origin of the specific Feynman rules for vacuum diagrams: (i)
the last time integration just gives an extra factor $-i\beta$, and
(ii) one of the vertices must be kept fixed to type $1$ or type $2$. A
justification of these additional rules is given in \cite{Gelis1} and
won't be reproduced in the present paper.

The second situation in which a contribution of the vertical part is
left at the end of the calculation is encountered for self-energy
insertions between propagators. Indeed, in that case, the frequency
$\Sigma$ can be the difference $\omega_{{\imb k}_1}-\omega_{{\imb
    k}_2}$ of the incoming and outgoing on-shell energies while
3-momentum conservation imposes ${\imb k}_1={\imb k}_2$, i.e.
$\Sigma=0$. This is the situation I will study in detail in the next
section. It is worth noticing that compared to \cite{Gelis1}, the
insertion of self-energies is shown to be the only situation in which
the vertical part contributes.\footnote{In \cite{Gelis1}, I identified
  the condition $\Sigma=0$ as the necessary condition to have a
  contribution of the vertical part in $I_{\cal C}(\Sigma)$, but
  didn't realize that this condition is relevant only if it defines a
  sub-manifold of strictly positive measure.}

\section{Effect of the vertical part\\
 on the RTF Feynman rules}
\label{sec:effect}
\subsection{Basic example}
I now study in detail the case of self-energy insertions which is the
only one in which the self-energy contributes, besides vacuum
diagrams, in order to show that the contribution of ${\cal C}_v$ is
automatically included by the matrix formalism provided that one uses
$|k^o|$ for the argument of statistical weights.  The general
philosophy of the proof is to start from a Green's function expressed
in time coordinates, for which we have unambiguous Feynman rules.
Then we have to Fourier transform it in order to obtain the
corresponding matrix. Finally, we must deduce from the result the
Feynman rules in Fourier space that would have given the same
function.  I will start by the trivial example of repeated mass
insertions. But contrary to \cite{Gelis1} where this example was only
used as an illustration for the more general case of self-energy
insertions, this example is in the present paper at the very heart of
the proof. Indeed, I show in the next paragraph that the most general
case can be reduced to the trivial one by making use of simple
algebraic properties of the contour integration.

The object we are interested in is the propagator obtained after the
resummation of an additional mass term $-i\mu^2$:
\begin{equation}
G(x^o_1,x^o_2)\equiv \sum\limits_{n=0}^{+\infty}(-i\mu^2)^n
(G_o*\cdots*G_o)(x_1^o,x_2^o)\; ,
\end{equation}
where the convolution product appearing at order $n$ in the sum
contains $n+1$ factors. Of course, the result of this sum is well
known without the need of performing the calculation:\footnote{For the
  term of order $n$ in the infinite sum, this result implies:
\begin{equation}
G_o*\cdots*G_o={1\over{n!}}
\left[i{{\partial}\over{\partial m^2}}\right]^n G_o\; ,
\end{equation}
a relation known as the mass derivative formula \cite{FujimMUO1}.  }
\begin{equation}
G(x^o_1,x^o_2)=\left.G_o(x^o_1,x^o_2)\right|_{m^2\to m^2+\mu^2}\; ,
\label{eq:mass-shift}
\end{equation}
where the notation $m^2\to m^2+\mu^2$ means that each occurrence of
$m^2$ in $G_o$ is replaced by $m^2+\mu^2$.  Since the Fourier
transform given by Eq.~(\ref{eq:FT-def}) does not involve the mass,
the above result for the resummed propagator also holds for its
Fourier transform. Therefore, we have to find out the Feynman rules
that would give the matrix propagator in which the mass squared is
translated by an amount equal to $\mu^2$. Let us now do the same
resummation in the matrix formalism by making use of
Eq.~(\ref{eq:feyn-naive}), in order to determine how it should be
modified in order to reach the expected result
Eq.~(\ref{eq:mass-shift}). To that effect, it is convenient to
factorize the free matrix propagator as follows
\begin{equation}
G_o(k)=U(k)\pmatrix{&\Delta_{_{F}}(k)&0\cr
&0&\Delta_{_{F}}^*(k)\cr}U(k)\; ,
\end{equation}
where $\Delta_{_{F}}\equiv i{\mathbbm P}/(k^2-m^2)+\pi\delta(k^2-m^2)$
is the usual Feynman propagator, and $U(k)$ is a matrix containing
the statistical factors:
\begin{equation}
U(k)=\pmatrix{
&\sqrt{1+n_{_{B}}}
&(\theta(-k^o)+n_{_{B}})/\sqrt{1+n_{_{B}}}\cr
&(\theta(k^o)+n_{_{B}})/\sqrt{1+n_{_{B}}}
&\sqrt{1+n_{_{B}}}\cr
}\; .
\end{equation}
At this stage, it seems that we still have the choice ($|k^o|$ or
$\omega_{\imb k}$) for the arguments of the statistical weights. In the
matrix formalism, the resummation is performed by
\begin{eqnarray}
G(k)&&=\sum\limits_{n=0}^{+\infty}(-i\mu^2)^n G_o(k)[\tau_3 G_o(k)]^n
\nonumber\\
&&=U(k)\left[\sum\limits_{n=0}^{+\infty}(-i\mu^2)^n 
D_o(k)[\tau_3D_o(k)]^n \right]U(k)
\nonumber\\
&&=U(k)\left[\left.D_o(k)\right|_{m^2\to m^2+\mu^2}\right]U(k)\; ,
\end{eqnarray}
where I denote $D_o(k)\equiv{\rm
  Diag}(\Delta_{_{F}}(k),\Delta_{_{F}}^*(k))$.  In order to do the
sum, I have used the algebraic relation $U\tau_3 U=\tau_3$.  It is now
obvious that if we want to have the relation
$G(k)=\left.G_o(k)\right|_{m^2\to m^2+\mu^2}$, we need
$U(k)=\left.U(k)\right|_{m^2\to m^2+\mu^2}$, which means that the
matrix $U(k)$ should be independent of $m^2$. The only way to achieve
that is to use $|k^o|$ as the argument of $n_{_{B}}$ in $U(k)$.

Therefore, we have justified in the case of this simple example the
fact that the prescription $n_{_{B}}(|k^o|)$ should be used in the RTF
Feynman rules in order to get the correct result. Moreover, since we
have seen in section \ref{sec:origin} that the vertical part enables
to take into account the interaction (here the term $-i\mu^2$) in the
density operator, i.e. in the statistical factors, we can conclude
that choosing the right argument for the statistical functions
reintroduces the contribution of the vertical part in the result.

\subsection{Repeated self-energy insertions}
\begin{figure}[htbp]
\centerline{\resizebox*{6cm}{!}{\includegraphics{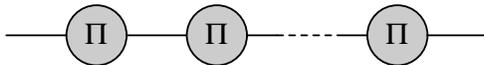}}}
\caption{Generic configuration giving a contribution of the
  vertical part.\label{fig:vert-contr}}
\end{figure}

I want now to generalize the previous result concerning the
$n_{_{B}}(|k^o|)$ prescription to the general case of self-energy
    insertions, illustrated on figure \ref{fig:vert-contr}.
This situation seems more complicated at first sight since we don't
know {\it a priori\/} the result. The calculation to be performed in
time coordinates is
\begin{equation}
G(x_1^o,x_2^o)\equiv\sum\limits_{n=0}^{+\infty}
(G_o*\Pi*\cdots*\Pi*G_o)(x_1^o,x_2^o)\; ,
\end{equation}
where the term of order $n$ in the right hand side contains $n$
factors $\Pi$ and $n+1$ factors $G_o$. This is where the properties of
the contour convolution discussed in appendix \ref{sec:appB} are quite
helpful. Indeed, if we use now the commutativity of this product of
convolution (which holds here since all the convoluted objects satisfy
KMS), we can rewrite
\begin{equation}
G(x_1^o,x_2^o)=\sum\limits_{n=0}^{+\infty}
((G_o*\cdots*G_o)*(\Pi*\cdots*\Pi))(x_1^o,x_2^o)\; .
\end{equation}
Now that the free propagators $G_o$ are grouped together, we have
reduced the problem to the previous one. Indeed, we know that we
don't have any contribution of the vertical part in the convolution of
two objects if at least one of them is one particle irreducible, which
is the case of $\Pi$. Therefore, ${\cal C}_v$ does not contribute in
the product $\Pi*\cdots*\Pi$, and we can obtain its Fourier transform
with the Feynman rules (Eq.~(\ref{eq:feyn-naive})) established under
the hypothesis that ${\cal C}_v$ does not contribute. The only problem
related to ${\cal C}_v$ comes from the product $G_o*\cdots*G_o$ which
has already been considered in the previous paragraph. Its Fourier
transform is obtained by the Feynman rules with the
$n_{_{B}}(|k^o|)$ prescription. Therefore, the Fourier transform is
given by
\begin{equation}
G(k)=\sum\limits_{n=0}^{+\infty}
G_o(k)[\tau_3 G_o(k)]^n \Pi(k)[\tau_3 \Pi(k)]^{n-1}\; ,
\end{equation}
in which one should use the $n_{_{B}}(|k^o|)$ prescription for the
$n+1$ $G_o$'s. At this stage, it is trivial to put the various factors
back into a more natural order to get\footnote{This is possible
  because we can write $G_o(k)=U D_o U$ and $\Pi(k)=U P U$ with $D_o$
  and $P$ diagonal matrices, and because $U$ satisfies $U\tau_3
  U=\tau_3$. This merely says that the commutativity of the contour
  convolution is transported in the matrix formalism.}
\begin{equation}
G(k)=\sum\limits_{n=0}^{+\infty}
G_o(k)[\tau_3 \Pi(k)\tau_3 G_o(k)]^n\; .
\end{equation}

This trick based on the commutativity of the contour convolution
enabled us to reduce the general case to the simpler one treated in
the previous paragraph, and to see again that the argument of the
statistical weights for the propagators along the chain\footnote{For
  the other propagators, the prescription for the statistical factors
  is indifferent.} must be $|k^o|$.

\section{Concluding remarks}
In this paper, I have given a new, quite compact, justification for
the matrix formalism for the RTF in Fourier space. The focus has been
on a correct treatment of the vertical part of the time path.  In
particular, no use is made of the limits $t_{_{I}}\to-\infty$ and
$t_{_{F}}\to+\infty$, since KMS implies a total independence of the
Green's functions with respect to these parameters. The justification
is made in three steps: (i) identify the diagrams in which the
vertical part contributes (ii) show in the case of mass insertions
that the contribution of ${\cal C}_v$ is included by the $n(|k^o|)$
prescription and (iii) use simple properties of the contour
convolution to reduce the general case to the previous one.

The present justification is complementary to that of Le Bellac and
Mabilat, since it provides a better control on which are the
topologies receiving a contribution from the vertical part, while in
\cite{BellaM1,Mabil1} all the topologies appear on the same footing.
Compared to that of \cite{Gelis1}, this proof is more complete since
the situations in which the vertical part contributes are clearly
delimited, and the end of the proof is considerably simplified by
making use of the commutativity of the contour convolution.

I would also like to emphasize again on the physics encoded in the
vertical part of the path. Indeed, since we know that the role of the
vertical part in the perturbative expansion is to extract the
dependence upon the coupling constant contained in the density
operator, it was obvious right from the beginning that its effects on
the Feynman rules could only affect the statistical factors.

To end this paper, it is worth making a comment on the Keldysh
formalism \cite{Keldy1} used in out-of-equilibrium situations. This
formalism is based on a time path which does not contain the vertical
part ${\cal C}_v$. Indeed, there is no need for it here since the
initial density operator is not related to the Hamiltonian and
therefore does not contain the coupling constant. All the properties
of the equilibrium Green's functions that are related to the presence
of the vertical part are lost: out-of-equilibrium Green's functions
depend explicitly on the initial time $t_{_{I}}$, and are not
invariant under time translation. For this reason, going to Fourier
space is also much less straightforward.

\appendix

\section{Path independence of contour integrations}
\label{sec:appA}
For the purpose of discussing the effect of the vertical part of the
path in the real time formalism, we need first to recall some basic
properties of the contour integration.

The most noticeable property of this integration is that it gives a
result which is independent of the initial time $t_{_{I}}$ used to
define the path \cite{Gelis1}. This property is in fact a quite direct
consequence of the KMS relations satisfied by propagators appearing in
the perturbative expansion, and was to be expected given the physical
meaning of thermal equilibrium. 

Indeed, we can write any Green's function $G(x_1^o,\cdots,x^o_n)$
calculated perturbatively as:
\begin{equation}
G(x_1^o,\cdots,x^o_n)\equiv \left[\prod\limits_{i=1}^{V}
\int_{{\cal C}}dy_i^o\right] \; g(x_1^o,\cdots,x^o_n|y_1^o,
\cdots,y^o_{_{V}})\; ,
\end{equation}
where $V$ is the total number of vertices in the diagram, and the
$y_i^o$ the inner times. Now, if we consider a function 
\begin{equation}
a(y^o)\equiv\theta_c(y^o-y_+^o)a^+(y^o)+\theta_c(y^o_--y^o)a^-(y^o)
\end{equation}
on the path ${\cal C}$, with holomorphic functions $a^{\pm}(y^o)$, and
then calculate the integral
\begin{equation}
A\equiv\int_{{\cal C}}dy^o\;a(y^o)\; ,
\end{equation}
we have the following two properties:

\noindent (i) $A$ depends only on the extremities of the path, 
and on the other times $y_\pm^o$, but not on its precise shape.

\noindent (ii) we have $dA/dt_{_{I}}=a(t_{_{I}}-i\beta)-a(t_{_{I}})$.

Looking now at the structure of the bare propagators (see
Eq.~(\ref{eq:bare-prop})), we see that the integrand $g$ satisfies the
conditions of the previous lemma with the additional property of
taking the same value at both extremities of the path for each inner
variable $y^o_i$. Applying therefore (ii), we conclude that the
function $G(x_1^o,\cdots,x^o_n)$ is independent upon $t_{_{I}}$. Using
then the possibility to deform the path (i), we can change $t_{_{F}}$
without changing the result of the integrals. The function
$G(x_1^o,\cdots,x^o_n)$ is therefore also independent of $t_{_{F}}$.
Finally, the only dependence of $G(x_1^o,\cdots,x^o_n)$ upon the path
comes through the external times $x_i^o$ which are supposed to be on
the path.

\section{Properties of the contour convolution}
\label{sec:appB}
In order to deal simply with self-energy insertions, it is convenient
to discuss first a few properties of the contour convolution defined
by Eq.~(\ref{eq:conv-def}).

The first obvious property is that the result is independent of both
$t_{_{I}}$ and $t_{_{F}}$ provided that the two functions one is
convoluting satisfy the KMS relations and correspond to two particles
of the same nature.\footnote{This limitation is not important in
  practice since convoluting a bosonic function with a fermionic one
  would be totally meaningless.}

In order to simplify the study of this operation for functions
satisfying KMS, the first step is to write two-point functions by
means of their spectral representation:\footnote{It is possible to
  group the two terms in this sum in order to obtain a single term
  containing the full free propagator. This splitting is natural here
  since $G_{o,s}^E$ is the smallest part of the free propagator that
  still satisfies KMS. Any property that is a consequence of KMS can
  be obtained by limiting the study to this very simple piece.}
\begin{equation}
F(x^o_1,x^o_2)=\sum\limits_{s=\pm}\int_{0}^{+\infty}dE\,f_s(E)
\,G_{o,s}^E(x^o_1,x^o_2)\; ,
\end{equation}
where the $G_{o,s}^E$ are the building blocks of the free propagator
given by Eq.~(\ref{eq:G-o-E}). If one uses this spectral
representation, it is sufficient to limit the study of the operation
$*$ to its action on simple objects like $G_{o,s}^E$.

An elementary integration based on Eq.~(\ref{eq:int-parts}) gives
immediately:
\begin{equation}
G_{o,\epsilon}^A*G_{o,\eta}^B={1\over{i(\epsilon A-\eta B)}}
\left[
\epsilon G^B_{o,\eta}-\eta G_{o,\epsilon}^A
\right]\; .
\end{equation}
We notice that the result is unchanged if we permute the two objects
we are convoluting. This property is transported to general two-point
functions through their spectral representation: the contour
convolution is commutative.

Iterating the above relation, we obtain:
\begin{eqnarray}
\left(G_{o,\epsilon}^A*G_{o,\eta}^B\right)*G_{o,\mu}^C&&
={{\eta}\over{i(\epsilon A-\eta B)}}{{\mu}\over{i(\epsilon A- \mu C)}}
G^A_{o,\epsilon}\nonumber\\
&&+{\epsilon\over{i(\eta B-\epsilon A)}}
{\mu\over{i(\eta B-\mu C)}}
G_{o,\eta}^B\nonumber\\
&&+{\epsilon\over{i(\mu C-\epsilon A)}}
{\eta\over{i(\mu C-\eta B)}}
G_{o,\mu}^C\; .
\end{eqnarray}
The remarkable property of this result is its symmetry under any
permutation of the three objects one is convoluting. Again, this is
trivially extended to any triplet of two-point functions: the contour
convolution is associative.

To conclude this appendix, one can say that as far as functions
satisfying KMS are concerned, the contour convolution possesses the
same basic properties as the ordinary convolution product.


\begin{thebibliography}{10}

\bibitem{KobesK1}
{R.L. Kobes, K.L. Kowalski}, Phys. Rev. {\bf D} {\bf 34}, 513 ({1986}).

\bibitem{KobesSW1}
{R.L. Kobes, G.W. Semenoff, N. Weiss}, Z. Phys. {\bf C} {\bf 29}, 371 ({1985}).

\bibitem{LandsW1}
{N.P. Landsman, Ch.G. van Weert}, Phys. Rep. {\bf 145}, 141 ({1987}).

\bibitem{FurnsS1}
{R.J. Furnstahl, B.D. Serot}, Phys. Rev. {\bf C} {\bf 44}, 2141 ({1991}).

\bibitem{Marin1}
{M. Marinaro}, Phys. Rep. {\bf 137}, 81 ({1986}).

\bibitem{NiemiS1}
{A.J. Niemi, G.W. Semenoff}, Nucl. Phys. {\bf B} {\bf 230}, 181 ({1984}).

\bibitem{Niega1}
{A. Niegawa}, Phys. Rev. {\bf D} {\bf 40}, 1199 ({1989}).

\bibitem{Evans6}
{T.S. Evans}, Phys. Rev. {\bf D} {\bf 47}, 4196 ({1993}).

\bibitem{EvansP1}
{T.S. Evans, A.C. Pearson}, Phys. Rev. {\bf D} {\bf 52}, 4652 ({1995}).

\bibitem{Gelis1}
{F. Gelis}, Z. Phys. {\bf C} {\bf 70}, 321 ({1996}).

\bibitem{BellaM1}
{M. Le Bellac, H. Mabilat}, Phys. Lett. {\bf B} {\bf 381}, 262 ({1996}).

\bibitem{Mabil1}
{H. Mabilat}, Z. Phys. {\bf C} {\bf 75}, 155 ({1997}).

\bibitem{Evans1}
{T.S. Evans}, Z. Phys. {\bf C} {\bf 36}, 153 ({1987}).

\bibitem{Evans2}
{T.S. Evans}, Z. Phys. {\bf C} {\bf 41},333 ({1988}).

\bibitem{RammeS1}
{J. Rammer, H. Smith}, Rev. of Modern Physics {\bf 58}, 323 ({1986}).

\bibitem{FujimMUO1}
{Y. Fujimoto, H. Matsumoto, H. Umezawa, I. Ojima}, Phys. Rev. {\bf D} {\bf 30},
  1400 ({1984}).

\bibitem{Keldy1}
{L.V. Keldysh}, Sov. Phys. JETP {\bf 20}, 1018 ({1964}).

\end{thebibliography}
\end{document}